\documentclass[12pt,epsf,epsfig,psfig]{article}
\usepackage{graphicx}
\usepackage{epsfig}
\usepackage{cite}
\def\J{$J/\psi$}
\def\j{J/\psi}

\def\P{$\psi'$}

\def\U{$\Upsilon$}

\def\C{c{\bar c}}

\def\P{$\psi'$}

\def\U{$\Upsilon$}

\def\C{c{\bar c}}

\def\NP{{ Nucl.\ Phys.\ }}
\def\PL{{ Phys.\ Lett.\ }}
\def\PR{{ Phys.\ Rev.\ }}

\def\PRL{{ Phys.\ Rev.\ Lett.\ }}

\def\ZP{{ Z.\ Phys.\ }}
\def\EPJ{{Eur.\ Phys.\ J.\ }}

\def\NP{{ Nucl.\ Phys.\ }}
\def\PL{{ Phys.\ Lett.\ }}
\def\PR{{ Phys.\ Rev.\ }}

\def\PRL{{ Phys.\ Rev.\ Lett.\ }}

\def\ZP{{ Z.\ Phys.\ }}
\def\EPJ{{Eur.\ Phys.\ J.\ }}

\def\be{\begin{equation}}
\def\ee{\end{equation}}
\def\lsim{\raise0.3ex\hbox{$<$\kern-0.75em\raise-1.1ex\hbox{$\sim$}}}
\def\gsim{\raise0.3ex\hbox{$>$\kern-0.75em\raise-1.1ex\hbox{$\sim$}}}

\begin{document}

\parindent=0pt 

{\small April 2012} \hfill {\small BI-TP 2013/06}

\vskip0.9cm

\centerline{{\Large \bf Calibrating the In-Medium Behavior of Quarkonia}}

\vskip0.5cm

\centerline{\bf Helmut Satz}

\bigskip

\centerline{Fakult\"at f\"ur Physik, Universit\"at Bielefeld}

\centerline{D-33501 Bielefeld, Germany}

\medskip

\centerline{satz@physik.uni-bielefeld.de}

\vskip0.5cm

\centerline{\bf Abstract}

\medskip

Quarkonium production has been considered as a tool to study the 
medium formed in high energy nuclear collisions, assuming that the 
formation of a hot and dense environment modifies the production pattern 
observed in elementary collisions. The basic
features measured there are the relative fractions of hidden to
open heavy flavor and the relative fractions of the different hidden
heavy flavor states. Hence the essential question is if and how these
quantities are modified in nuclear collisions. We show how the relevant 
data must be calibrated, i.e., what reference has to be used, in order to 
determine this in a model-independent way.

\vskip1cm

\begin{figure}[htb]
\centerline{\epsfig{file=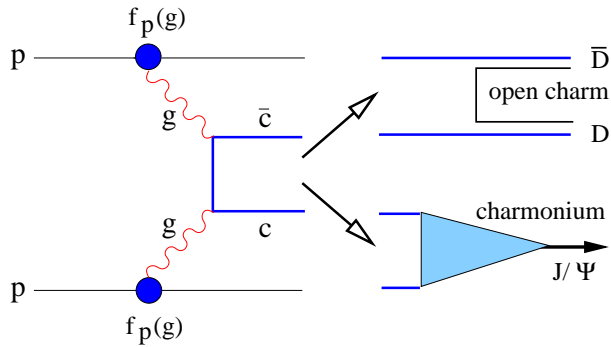,width=8cm}}
\caption{Schematic view of \J~production in $pp$ collisions}
\label{pp}
\end{figure}

The original suggestion for charmonium production as a means to 
test the formation of a deconfined medium in high energy nuclear
collisions was based on the idea that color screening in such a 
medium would prevent the binding of charm quarks to a color neutral 
\J~\cite{MS}. The \J~production process in elementary
hadronic collisions (taking $pp$ as example) begins
with the formation of a $\C$ pair; this pair can then either lead to
open charm production (about 90 \%) or subsequently bind to form
a charmonium state (about 10 \% for all charmonia). A schematic 
illutration (Fig. \ref{pp}) shows the dominant high energy reaction 
through gluon fusion. 

The initial $\C$ production can be calculated in terms of the parton 
distribution functions $f_p$ of the relevant hadrons and the pertubative
partonic cross section. The full description of charmonium binding has 
so far resisted various theoretical attempts; on the other hand, the
process is in good approximation independent of the incident hadronic
collision energy \cite{Gavai,HP}. This is a consequence of the fact that 
the heavy quark propagator in the reaction $gg \to \C$ strongly dampens 
the mass variation of the $\C$ pair with incident energy. Thus the
fractions of the produced $\C$ system into hidden vs.\ open charm
as well as those for the different charmonium states are approximately
constant; once determined at one energy, they remain the same also for
different energies. As a result, the phenomenological color evaporation 
model \cite{CE1,CE2,CE3,CE4} provides a good description of charmonium 
production 
through the form
\be
\sigma_{hh\to \j}(s) = g_{\C \to \j}~\!\sigma_{hh \to \C}(s),
\label{cem}
\ee
and correspondingly for the other charmonium states. Here
the constant $g_{\C \to \j}$ specifies what fraction of the total
$\C$ production cross section goes into \J~production; in $pp$
collisions it is typically about 2 \%. The set of the different
constants $g_{\C \to i}$ for the different charmonium states $i$
thus effectively characterizes charmonium production in the absence
of a medium. 

\medskip

A further important aspect of quarkonium production in elementary
collisions is that the observed
(1S) ground states \J~and \U~are in both cases partially produced through
feed-down from higher excited states \cite{FD1,FD2,FD3,FD4}.
Of the observed \J~rates, only some
60\ \% is a directly produced $\j(1S)$ state; about 30 \% comes from 
$\chi_c(1P)$ and 10 \% from $\psi'(2S)$ decay. 
Because of the narrow width of the excited states, 
their decay occurs well outside any interaction region.

\medskip

The features we have here summarized for charm and charmonium production
are readily extended to that of bottom and bottomonium. To simplify
the discussion, we shall continue referring to the charmonium case,
keeping in mind that all arguments apply as well to bottomonia. Given 
the patterns observed in elementary collisions, 
we want to see how they are modified 
in the presence of a medium, as provided by nuclear collisions. From the
point of view of production dynamics, one way 
such modifications can arise is as 
{\sl initial state effects}, which take place before
the $\C$ pair is produced. The main possibilities considered so far are
nuclear modifications of the parton distribution functions (shadowing or
antishadowing) and a possible energy loss of the partons passing through
the nuclear medium to produce the $\C$. Once produced, the pair can encounter 
{\sl final state effects}, either in the form of a phase space shift 
already of the $\C$, e.g., through an energy loss of the unbound
charm quarks, or through effects on the nascent or fully formed
charmonium state.  Such effects may arise from the passage through the
cold nuclear medium, or because of the presence of the medium newly
produced in the nuclear collision. The latter is evidently what we have
in mind when we want to use quarkonia to study quark-gluon plasma
production. The difficulties encountered over the past 
years \cite{Klu1,Klu2} 
in arriving at a conclusive analysis of the relevant nuclear collision 
data on \J~production are largely due to the 
problem of parametrizing the different effects and then constructing a
convincing model correctly incorporating all of them. We want to show here 
that today experimental means have become available which allow us to 
calibrate the measured results in a way which avoids these difficulties.

\medskip

\begin{figure}[htb]
\centerline{\epsfig{file=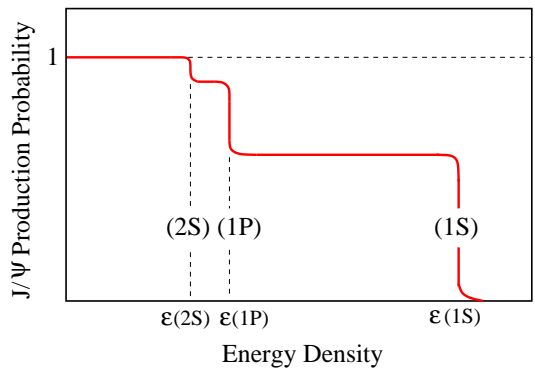,width=7cm}}
\caption{Schematic view of sequential \J~suppression in a deconfined medium}
\label{seq}
\end{figure}

Before we turn to the problem of determining medium effects on quarkonium
production, we recall the two main conceptual approaches on what may happen;
the two can perhaps
best be labelled {\sl suppression} and {\sl enhancement}. Color screening
in a quark-gluon plasma will decrease the quarkonium binding, both in
strength and in its spatial range, and this should for sufficiently
energetic nucleus-nucleus collisions lead to quarkonium dissociation or
melting. Since the larger and less tightly bound states will melt at
lower temperature or energy density than the ground states, color
screening will produce {\sl sequential suppression} \cite{Seq1,Seq2}. We 
illustrate this
for the \J. After an initial threshold melting the \P~and hence remo\-ving
its feed-down component for \J~production, there will be a second threshold
for $\chi_c$ melting and then finally a third, at which the direct 
$\j(1s)$ is dissociated. The resulting pattern is 
illustrated in Fig.\ \ref{seq}. 
We have here introduced something denoted as \J~survival probability.
Theoretically, this is the chance of a \J~to persist as a bound state 
in a deconfined medium. How to properly
define this quantity as a useful observable 
in experimental studies is the subject of this paper.

\medskip

\begin{figure}[htb]
\centerline{\epsfig{file=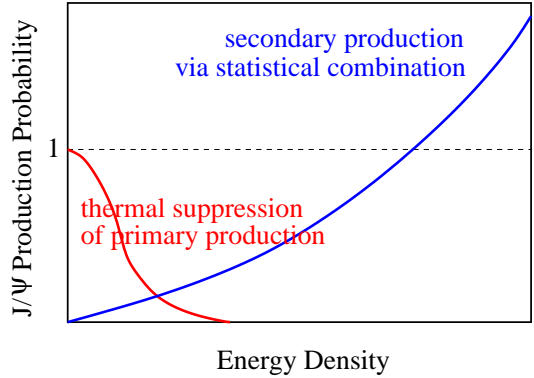,width=7cm}}
\caption{Schematic view of secondary production through statistical
combination}
\label{stat}
\end{figure}

The other alternative, \J~enhancement, assumes, in accord with color
screening and the resulting suppression, that at sufficiently
high energies there is an almost complete dissociation of the \J's
produced in primary nucleon-nucleon interactions. On the other hand, 
at such collision energies, these interactions lead to abundant
$\C$ production; the rate for this process grows faster than that for 
the production of light quarks, and if the $\C$ pairs remain present in 
the evolution of the medium, the system will at the hadronisation point
show an oversaturation of charm, compared to the 
predicted thermal abundance. 
If these charm quarks have become part
of an equilibrated medium and as such undergo hadronisation in the form
of statistical combination, then such secondary charmonium formation
can convert more $\C$ pairs into \J's than the dynamical primary production
mechanism, thus leading to an effective \J~enhancement 
\cite{SR1,SR2,SR3,SR4}, as shown in 
Fig.\ \ref{stat}. We note that this scenario invokes a new and so far
unknown binding dynamics to form charmonia; the outcome is assumed to be
determined simply by the relative abundance of charm quarks \cite{PBM2}.  

\medskip

We now come to our central question: how to calibrate the \J~survival?
Since we are interested in using quarkonium production as a tool to
study the medium produced in nuclear collisions, our primary concern
is not if such collisions produce more or fewer $\C$ pairs than 
proton-proton collisions, but rather if the presence of
the medium modifies the fraction of produced $\C$ pairs going into
charmonium formation. In other words, the crucial quantitity is the 
amount of charmonium production relative to that of open charm
\cite{Sridhar}. To 
illustrate: in $pp$ collisions, about 2 \% of the total $\C$ production
goes into \J's. If in high energy nuclear collisions the total $\C$
production rate were reduced by a factor two, but we still have 2 \%
of these going into \J's, then evidently $AA$ collisions do not modify
\J~binding. This is, of course, strictly true only for fully integrated
rates, which is difficult to check; we return to the issue of
phase space variations shortly. But we note here already that in 
the example just given, the number of \J's produced in $AA$ collisions 
would be half of that obtained by scaling the results from $pp$ interactions. 
That is, however, not an indication of any \J~suppression; it is just
the consequence of having fewer $\C$ pairs to start with -- 
the formation probability of charmonia has remained the same as in $pp$ 
collisions, 2 \%. It is thus crucial to correctly calibrate the survival
in nuclear collisions. 

\medskip

To achieve that, we recall: both sequential suppression and
statistical enhancement state that the total number of $\C$ pairs
produced in nuclear collisions is distributed among hidden and open
charm differently than it is in proton-proton collisions. The essential
observable is thus the ratio of charmonium states to open charm in nuclear
collisions, compared to that in $pp$ interactions. Sequential suppression
predicts it to decrease with centrality and eventually vanish, 
statistical enhancement has it increasing with centrality, in many studies 
beyond the $pp$ value. Hence the relevant
observable is the fraction of charmonia to open charm, or more generally,
that of quarkonia to the relevant open heavy flavor production 
\cite{Sridhar,PBM1}. In this quantity, if measured over the entire phase 
space, the effects of possible initial state nuclear modifications -- 
shadowing/antishadowing, parton energy loss -- cancel out, so that
whatever changes it shows relative to the $pp$ pattern is due to final
state effects.  

\medskip

It should be noted here that in actual applications, it will most likely
not be necessary (nor generally possible) to measure the total open charm
production rate. The relative abundances of the different open charm states 
produced in high energy collisions are so far found to be in good agreement 
with the predictions of the statistical hadronization model \cite{BCMS,PBM3}, 
with a universal hadronisation temperature of some 150 - 170 MeV. For open 
charm production, this model assumes an initial dynamical (perturbative) 
$\C$ production; subsequently, at the hadronisation point, the charm quarks 
thus produced form open charm hadrons according to their statistical 
abundances. The relative production rates are thus totally determined in 
terms of the mass of the open charm state and the universal hadronisation 
temperature. Such a description is found to hold very well for open charm 
production in high energy interactions from $e^+e^-$ annihilation 
through $pp$ and $pA$ collisions; studies of the $AA$ behavior are underway. 
However, the model does not work for charmonium production, neither in
elementary ($e^+e^-$, $pp$) nor in nuclear interactions, since the binding 
mechanism forming charmonia is evidently of dynamical and not of statistical 
origin. In nuclear collisions, there are in addition medium effects on 
the binding.

\medskip

As a consequence, we expect that the {\sl relative abundances} of the 
different open charm states will not depend on the 
collision energy and remain essentially the same for elementary and nuclear 
collisions. Analogous to eq.\ (\ref{cem}), the production rate $N(hh \to D_i)$ 
for a specific open charm state $D_i$ is then a constant fraction of the 
total open charm rate $N(hh \to \C)$,
\be
N(hh \to D_i) \simeq {\bar g}_{\C \to D_i} N(hh \to \C).
\label{open}
\ee
The crucial observable, the rate of (\J)/open charm, can thus
in good approximation be taken as the rate of (\J)/$D_i$ for a specific 
$D$-meson state. This will, of course, greatly facilitate the analyis. 

\medskip

The final state effect of interest to us is that caused by the newly produced
medium. We therefore have to check to what extent the nascent or fully formed 
\J~is already dissociated by the cold nuclear matter of the target or the
projectile. This can be done by studying the ratio hidden to open charm in
high energy $pA$ collisions. It is in fact known that in such interactions
there exist reductions of \J~production beyond the scaled $pp$ results, but 
the origin of these is not unambiguously clarified. They could arise largely 
from initial state effects; but if there is a stronger suppression of 
\P~than of \J~at energies for which the state is formed inside the nuclear 
medium, this would indicate that there the interaction depends 
on its physical size. The different fates of the different quarkonium 
states in {\sl cold} 
nuclear matter are due to their different sizes, in a {\sl hot} medium 
due to the different binding energies. And dissociation by color screening 
should start with a threshold, while the break-up in nuclear matter is 
presumably continuous. Measurements of charmonia relative to open charm 
in $pA$ up to highest energies (RHIC, LHC) are therefore of great importance.

\medskip

\begin{figure}[htb]
\centerline{\epsfig{file=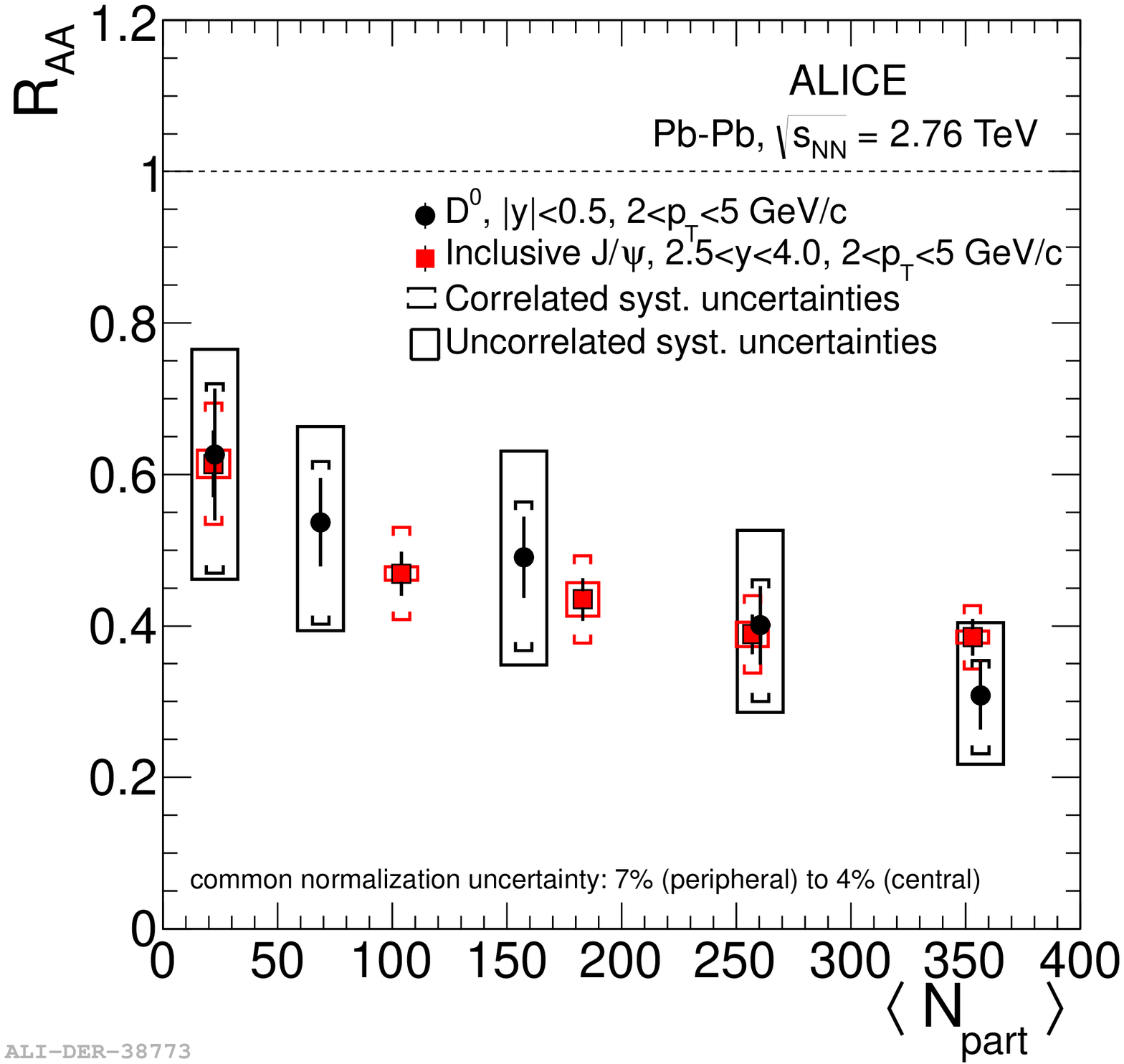,width=6.5cm}\hskip2cm
\epsfig{file=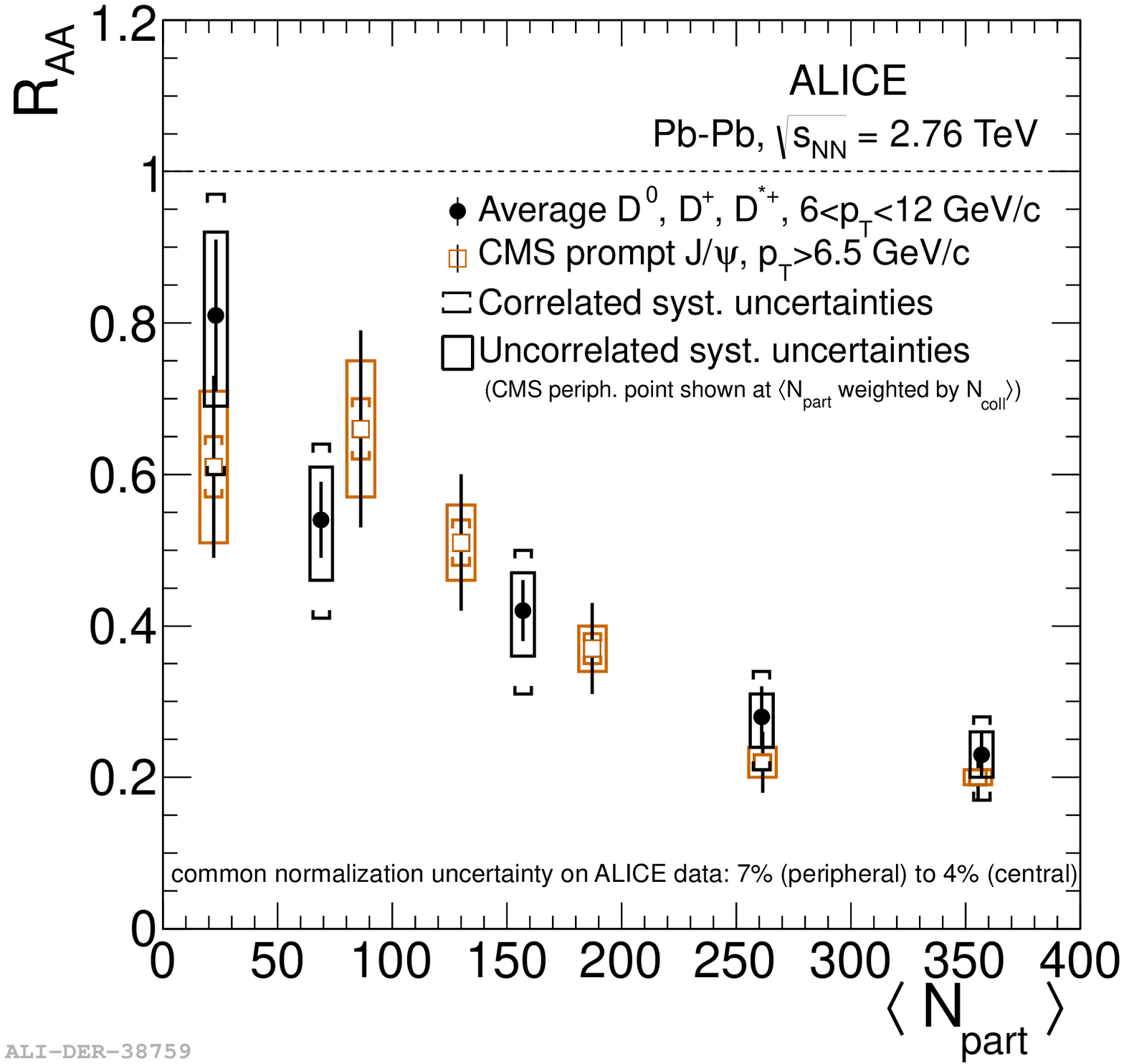,width=6.5cm}}
\hskip3.1cm (a) \hskip7.8cm (b)
\caption{LHC data from ALICE and CMS \cite{Alice1,Alice2,CMS},
comparing \J~production to open
charm production at intermediate (a) and high (b) transverse momenta}
\label{zaida}
\end{figure}

First applications of the in-medium charmonium study based on the
relative survival of charmonia vs.\ open charm were started last year,
using LHC data from ALICE and CMS \cite{Zaida,Alice1,Alice2,CMS}. In 
Fig.\ \ref{zaida}(a),
we show mid-rapidity ALICE data for \J~production at intermediate transverse 
momenta, compared to open charm production in a similar kinematic region.
In Fig.\ \ref{zaida}(b), the comparison is extended to larger transverse
momenta, using CMS data for \J~production. In both cases, \J~production 
relative to $pp$ results, scaled by the number of collisions, decreases with 
increasing centrality, as seen by the corresponding $R_{AA}$ values.

\medskip

\medskip

This decrease has at times been considered as suppressed \J~production. 
However, that is incorrect: the corresponding $R_{AA}$ for open charm 
production, as determined through $D$ measurements, shows within errors 
the same behavior. 
In other words, the reduction of the \J~is in complete 
agreement with that of open charm; there is neither suppression nor 
enhancement, the fraction of the produced $\C$ pairs going into 
\J~production has remained in the $AA$ collisions considered here the 
same as in the corresponding $pp$ interactions:
\be
R_{AA}(\j) = {N_{AA}(\j) \over n_c N_{pp}(\j)} = 
{N_{AA}(\C) \over n_c N_{pp}(\C)} = R_{AA}(\C),
\ee
with $n_c$ denoting the scaling factor for the number of collisions
at the corresponding centrality. We therefore have 
\be
{N_{AA}(\j) \over N_{AA}(\C)} = 
{N_{pp}(\j) \over N_{pp}(\C)} = g_{\C \to \j}.
\ee
at all centralities, in the kinematic regime indicated.
If we divide the \J~rates by the open charm rates in the same kinematic
region (large $P_T$), this ratio becomes centrality independent, and
if we normalize it to the corresponding value from $pp$ collisions,
it becomes unity. The correct and model-independent \J~survival probability
for experimental study is thus
\be
S_{\j} = \left({N_{AA}(\j) \over N_{AA}(\C)}\right)
/ \left({N_{pp}(\j) \over N_{pp}(\C)}\right) = {1 \over g_{\C \to \j}}
\left({N_{AA}(\j) \over N_{AA}(\C)}\right).
\ee
In the kinematic regime considered so far, it is indeed unity, there is
neither suppression nor enhancement.

\medskip

\begin{figure}[hbt]
\centerline{\epsfig{file=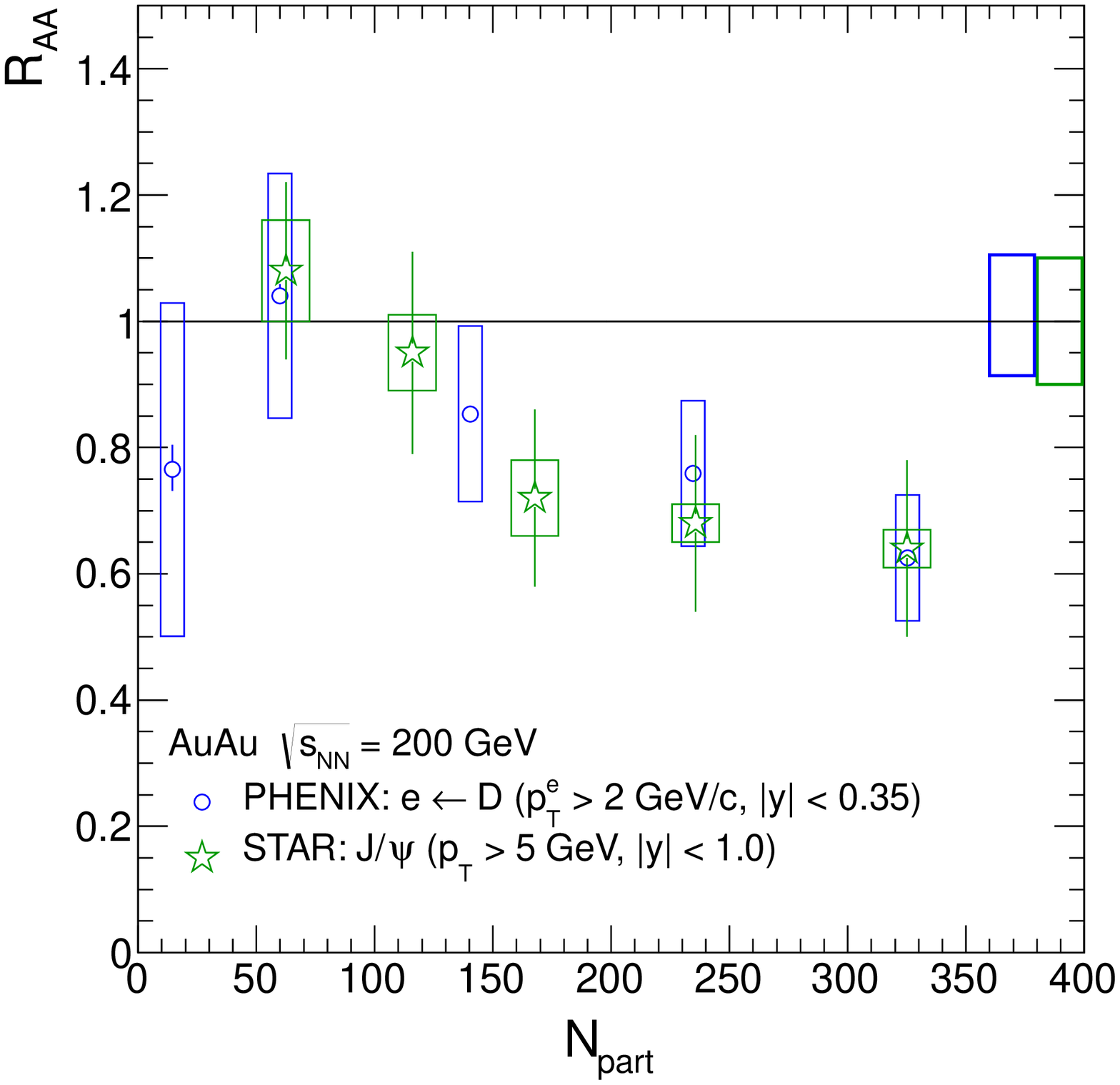,width=6.5cm}\hskip2cm
\epsfig{file=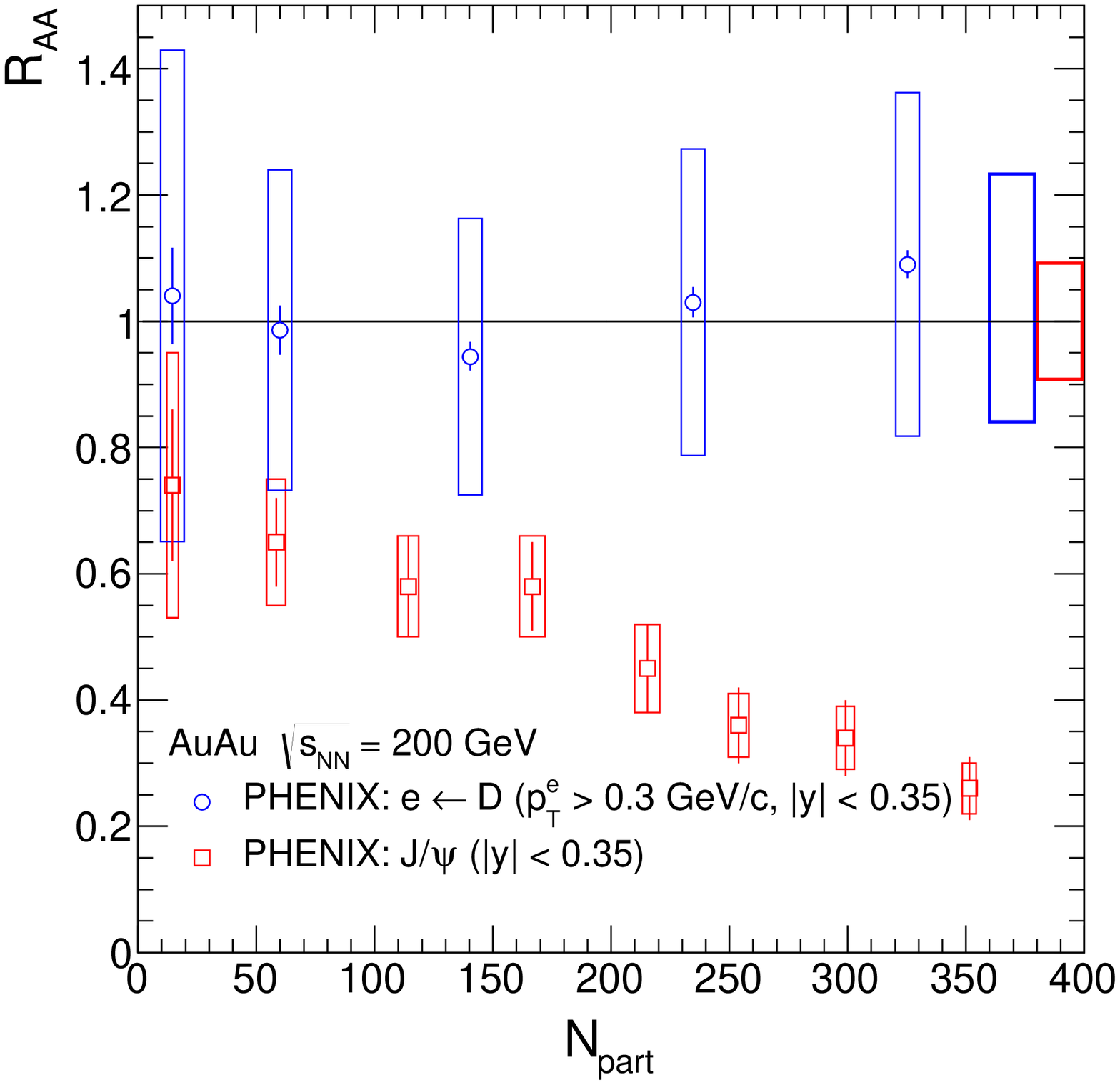,width=6.5cm}}
\vskip0.2cm

\hskip3.1cm (a) \hskip7.9cm (b)
\caption{RHIC data from PHENIX and STAR \cite{phenix1,phenix2,phenix3,star},
comparing \J~production to open
charm production at high (a) and low (b) transverse momenta}
\label{dahms}
\end{figure}

At the LHC, corresponding data for low $P_T$ open charm production is not 
yet available. At RHIC, however, it is provided by the PHENIX and STAR 
collaborations \cite{dahms,phenix1,phenix2,phenix3,star},
and the relevant comparisons of \J~vs.\ open charm production are shown
in Fig.\ \ref{dahms}. 
At high $P_T$ we have a similar behavior as at the LHC, no change of
\J~production relative to open charm. At low $P_T$, however,
the $R_{AA}(\C)$ of open charm is within errors
unity over the whole centrality range; in contrast, $R_{AA}(\j)$ 
decreases strongly and thus here gives the correct \J~survival probability. 
We see that 
now with increasing centrality, a smaller and smaller fraction of $\C$ pairs 
go into \J~production, with a suppression of up to 75\% for the most 
central collisions. The \J~production finally surviving could conceivably 
be largely due to corona interactions \cite{ramona}. 

\medskip

\begin{figure}[htb]
\centerline{\epsfig{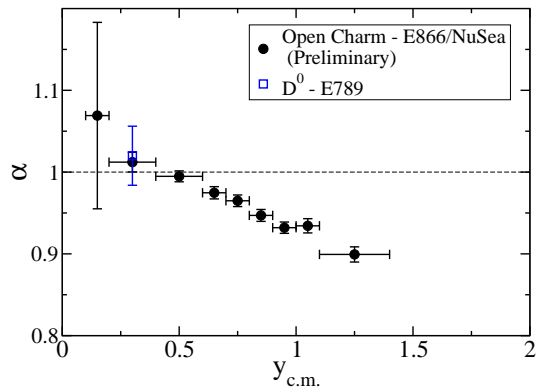}}
\caption{The rapidity dependence of open charm production in $pA$
collisions at 800 GeV, parametrized in the form $\sigma_{pA}=
A^{\alpha}\sigma_{pp}$ \cite{nusea}.}
\label{Mike}
\end{figure}

In this context it seems of interest to note that a decrease of the open
charm production rates between central and forward rapidity is observed
in both $pA$ and $AA$ collisions relative to $pp$ collisions,
at Fermilab (pA) and at RHIC (Cu-Cu).
We note in particular in the Fermilab data based on 800 GeV $pA$ collisions
(see Fig.\ \ref{Mike}) that at forward rapidity there is a suppression
of open charm production, relative to scaled $pp$ rates \cite{nusea}, while
at midrapidity there is not. A similar effect is observed in $Cu-Cu$
collisions with $\sqrt s = 200$ GeV at RHIC \cite{Cu}. If this is passed
on to charmonium production, one expects for RHIC and LHC data a smaller 
$R_{AA}(\j)$ at forward than at mid-rapidity, simply because there are
fewer $\C$ pairs there to form charmonia, no matter how.
Such an effect was indeed observed at RHIC and has often 
been considered a puzzle \cite{jpsirhic}.

\bigskip

The aim of this work was to show how quarkonium production has to be
calibrated in order to obtain experiment-based, model-independent
information on suppression
or enhancement due to the hot medium formed in nuclear collisions.
The data cited are given for illustration, to make the point.
In a more complete study,
the transverse momentum range of \J~and open charm should be correlated
more precisely. The crucial question now remains, of course, how the
open charm rates for low $P_T$ production at the LHC will behave.
That kinematic region is responsible for the bulk of $\C$ production,
and the issue to be decided is if nuclear collisions do or do not
lead to an overall reduction. At RHIC, that is not the case, as 
Fig.\ \ref{dahms}(b) shows.

\bigskip

\centerline{\bf \large Acknowledgements}

\medskip

It is a pleasure to thank Zaida Conesa del Valle, Torsten Dahms,
Louis Kluberg, Mike Leitch, Carlos Louren{\c c}o and J\"urgen Schukraft 
for helpful discussions. Particular thanks go to Torsten Dahms also
for pointing out the applicability of RHIC data for the analysis
and for providing me with the corresponding Figures.

\end{document}